# Tailoring spatiotemporal wavepackets via two-dimensional space-time duality

Wei Chen[1,#], An-Zhuo Yu[1,#], Zhou Zhou[2], Ling-Ling Ma[1,*], Ze-Yu Wang[1], Jia-Chen Yang[1], Cheng-Wei Qiu[2,*], Yan-Qing Lu[1,*]

[1]National Laboratory of Solid State Microstructures, Key Laboratory of Intelligent Optical Sensing and Manipulation, Collaborative Innovation Center of Advanced Microstructures, College of Engineering and Applied Sciences, Nanjing University, Nanjing 210023, China.

[2]Department of Electrical and Computer Engineering, National University of Singapore, Singapore 117583, Singapore.

[#]These authors contributed equally: Wei Chen, An-Zhuo Yu.

*Correspondence to: malingling@nju.edu.cn, chengwei.qiu@nus.edu.sg, yqlu@nju.edu.cn

**Abstract**
Space-time (ST) beams, ultrafast optical wavepackets with customized spatial and temporal characteristics, present a significant contrast to conventional spatial-structured light and hold the potential to revolutionize our understanding and manipulation of light. However, the progress in ST beam research has been constrained by the absence of a universal framework for their analysis and generation. Here, we introduce the concept of 'two-dimensional ST duality', establishing a foundational duality between spatial-structured light and ST beams. We show that breaking the exact balance between paraxial diffraction and narrow-band dispersion is crucial for guiding the dynamics of ST wavepackets. Leveraging this insight, we pioneer a versatile complex-amplitude modulation strategy, enabling the precise crafting of ST beams with an exceptional fidelity exceeding 97%. Furthermore, we uncover a new range of ST wavepackets by harnessing the exact one-to-one relationship between scalar spatial-structured light and ST beams. Our findings suggest a paradigm shift opportunity in ST beam research and may apply to a broader range of wave physics systems.

**Introduction**
Recent advances in quantum physics and optics have pushed the boundaries of our understanding of space-time (ST), a concept profoundly transformed since Einstein's theory of relativity challenged Newton's notion of absolute ST. One typical example of this progress is the theoretical prediction and experimental realization of time crystals, an extraordinary phase of matter that extends the concept of spatial periodicity into the time domain [1, 2]. In parallel, the realm of optics has uncovered temporal analogs to spatial phenomena through the study of paraxial diffraction of light beams and dispersion propagation of short optical pulses [3–14]. This concept of ST duality—first noticed six decades ago [3, 4] and later refined by Brian in 1994 through the development of temporal imaging theory [5, 6]—has inspired a plethora of research endeavors, such as temporal cloaking [7, 8], temporal reflection/refraction/diffraction [9–12], and temporal ghost imaging [13, 14]. Despite these advances, the challenge remains to integrate spatial and temporal dimensions for analysis within a single physical system or object.

    The exploration of two-dimensional (2D) ST beams has emerged as a promising approach to address this integration challenge. These optical wavepackets, which replace one spatial

dimension with time, serve as ST counterparts to spatial-structured light [15–27]. The shift to a 2D ST framework has endowed ST beams with unprecedented properties, including arbitrary group velocities [15–17], anomalous refraction [18], and manifestations of ST optical vortices (STOVs) exhibiting transverse orbital angular momentum (OAM) [20–27]. These emergent properties provide new possibilities for manipulating light and its interaction with matter.

Although the research into ST beams is progressing, driven by distinct motivations and interests, it follows a unique trajectory compared to the more mature field of conventional spatial-structured light. ST light sheets, for example, extend the range of 1D non-diffracting light beams with their continuous or discretized impulse ST spectra [19], representing specific trajectories where various tilted spectral planes intersect the free-space light-cone, with the plane's geometry determining their group velocity and refractive properties [19]. In contrast, research on STOVs centers around their transverse OAM [20–22], with evolution dynamics dominated by phase differences across their temporal frequency components during propagation [25, 26]. Generating STOVs involves introducing an ST helical phase, yet observed intensity profiles in recent experiments often deviate from ideal symmetrical distributions [26, 27]. The advancement of ST beams has been somewhat impeded by the lack of a universally accepted theoretical and experimental framework, which is essential for a more organized exploration and advancement of ST beams.

Here, we propose the concept of '2D ST duality' to bridge these gaps and unify the understanding and generation of ST wavepackets. Our framework not only highlights the unique properties of ST beams but also emphasizes their similarities with spatial-structured light. We show that synchronization or divergence in the behavior of ST wavepackets and spatial-structured light can be precisely controlled through material dispersion, rooted in the mathematical duality of the paraxial wave equations that govern these two beam types. This conceptual shift lays the foundation for an ST complex-amplitude modulation scheme that enables the generation of arbitrary ST beams with fidelity above 97%. Additionally, our framework establishes a precise one-to-one correspondence between ST beams and their spatial counterparts, leading to our discovery of a wealth of novel ST wavepackets. This work blurs the boundary between conventional structured light and ST beam research, and hints at broader implications for wave systems, such as acoustics and electron waves.

## Results
### Theory for 2D ST duality

It is well known that a light beam propagating in free space undergoes spatial spreading due to paraxial diffraction, whereas a short optical pulse experiences broadening when propagating in a dielectric due to material dispersion (see Figs. 1a and 1b). The concept of ST duality arises from the realization that these two processes are governed by a pair of mathematically equivalent diffusion equations [5, 6]:

$$\frac{\partial \psi(x;z)}{\partial z} = \frac{i}{2k_0} \frac{\partial^2 \psi(x;z)}{\partial x^2}, \tag{1}$$

$$\frac{\partial \psi(\tau;z)}{\partial z} = -\frac{i\beta_2}{2} \frac{\partial^2 \psi(\tau;z)}{\partial \tau^2}. \tag{2}$$

Here, $\psi(x;z)$ and $\psi(\tau;z)$ denote a monochromatic 1D spatial beam and a 1D optical pulse, respectively. For Eq. (1), $k_0 = \frac{\omega_0}{c}$ is the wavenumber, $\omega_0$ is the angular frequency, and $c$ is the speed of light in vacuum. In Eq. (2), $\tau = t - \frac{z}{v_g}$ is the local time in a pulse frame, $v_g$ is the group velocity, $\beta(\omega) \approx \frac{n(\omega)\omega}{c}$ is the propagation constant in the medium, $n(\omega)$ is the refractive index, $\beta_2 = d^2\beta(\omega)/d\omega^2|_{\omega=\omega_0}$ is the 2-order medium dispersion, and $\omega_0$ represents the center angular frequency. Performing separately spatial and temporal Fourier transforms $\tilde{\psi}(k_x; z) = \int dx\, \psi(x;z)\exp(-ik_x \cdot x)$ and $\tilde{\psi}(\Omega; z) = \int d\Omega\, \psi(\tau;z)\exp(i\Omega \cdot \tau)$ on Eqs. (1) and (2), we obtain $\tilde{\psi}(k_x; z) = \tilde{\psi}_x^0 \exp\left(-i\left(\frac{k_x^2}{2k_0}\right)z\right)$ and $\tilde{\psi}(\Omega; z) = \tilde{\psi}_\tau^0 \exp\left(i\left(\frac{\beta_2 \Omega^2}{2}\right)z\right)$, where $\tilde{\psi}(k_x; z)$ and $\tilde{\psi}(\Omega; z)$ are spatial and temporal Fourier spectra of the spatial light beam and optical pulse, $\tilde{\psi}_x^0 = \tilde{\psi}(k_x; z=0)$ and $\tilde{\psi}_\tau^0 = \tilde{\psi}(\Omega; z=0)$ are the initial spectra at $z=0$, and $\Omega = \omega - \omega_0$ is the detuning frequency. Obviously, the similar mathematical structures of the two phase factors $\exp\left(-i\left(\frac{k_x^2}{2k_0}\right)z\right)$ and $\exp\left(i\left(\frac{\beta_2 \Omega^2}{2}\right)z\right)$ underpin the ST duality, reflecting in various propagation phases for different spatial and temporal frequency components of the beam and pulse during their propagation (see Figs. 1a and 1b).

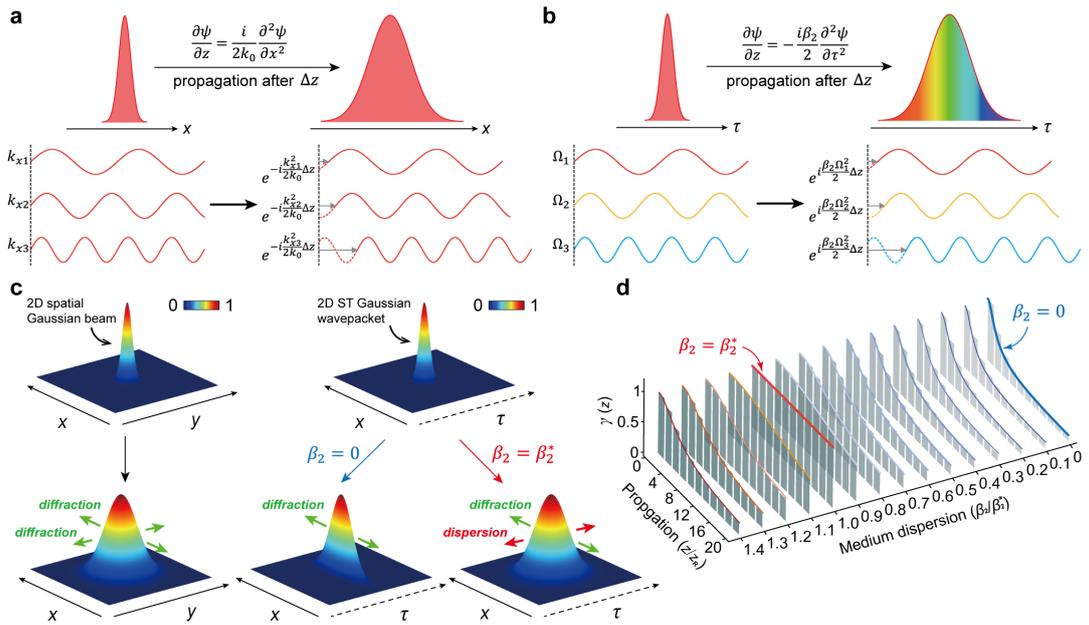

**Fig. 1 | From 1D to 2D ST duality. a**, The propagation of a monochromatic 1D spatial light beam, $\psi(x;z)$, over a distance $\Delta z$ in free space results in a broadening of its spatial width due to diffraction. This process can be understood as the accumulation of different propagation phases $\exp\left(-i\left(\frac{k_x^2}{2k_0}\right)\Delta z\right)$ by various spatial frequency components $k_x$ during propagation. **b**, Similarly, a 1D polychromatic short pulse, $\psi(\tau;z)$, upon propagating a distance $\Delta z$ in a dielectric with a 2-order dispersion of $\beta_2$, experiences a broadening of its pulse duration due to dispersion propagation. This can be understood as different temporal frequency components $\Omega$ accumulating various propagation phases $\exp\left(i\left(\frac{\beta_2 \Omega^2}{2}\right)\Delta z\right)$. Notably, the analogous mathematical structures of the propagation phases in **a** and **b**

reflects the underlying 1D ST duality. **c,** Concept of 2D ST duality. The propagation of a 2D ST Gaussian wavepacket $\psi(x,\tau;z)$ exhibits a duality symmetry with a 2D spatial Gaussian beam $\psi(x,y;z)$, when the material dispersion satisfies $\beta_2 = \beta_2^*$; otherwise, their behaviors diverge. **d,** The calculated similarity $\gamma(z)$ between a 2D ST Gaussian wavepacket propagating through a medium over 20 times the Rayleigh range $z_R$ with varying material dispersion, compared to a 2D spatial Gaussian beam propagating the same distance in free space. Note that the $\gamma(z)$ remains at unity ($\gamma = 1$) when $\beta_2 = \beta_2^*$; otherwise, $\gamma(z)$ decreases with propagation.

We next extend the concept of ST duality to a 2D framework, originated from the mathematical equivalence of the paraxial wave equations describing a monochromatic 2D spatial beam and an ST beam under the condition of anomalous dispersion, i.e., $\beta_2 < 0$: (see more details in Supplementary Text S1)

$$\frac{\partial \psi(x,y;z)}{\partial z} = \frac{i}{2k_0}\frac{\partial^2 \psi(x,y;z)}{\partial x^2} + \frac{i}{2k_0}\frac{\partial^2 \psi(x,y;z)}{\partial y^2}, \quad (3)$$

$$\frac{\partial \psi(x,\zeta;z)}{\partial z} = \frac{i}{2\beta_0}\frac{\partial^2 \psi(x,\zeta;z)}{\partial x^2} + \frac{i}{2\beta_0}\frac{\partial^2 \psi(x,\zeta;z)}{\partial \zeta^2}. \quad (4)$$

Here, $\psi(x,y;z)$ represents the monochromatic spatial beam and $\psi(x,\zeta;z)$ is the ST beam, with the propagation constant of the latter in a dielectric being $\beta_0 = n(\omega_0)k_0$. In Eq. (4), we introduce a length-scale parameter $\zeta = \frac{\tau}{\sqrt{-\beta_0\beta_2}}$, which quantifies the instantaneous spatial length scale of the ST beam in the $z$-direction. Since the spatial width of the ST beam in the $z$-direction also depends on the group velocity, i.e., $\zeta = v_g\tau$, we obtain the characteristic material dispersion $\beta_2^* = \frac{-c}{n(\omega_0)\omega_0 v_g^2}$ that satisfies the 2D ST duality symmetry. Moreover, despite the 2D ST duality indirectly implies a symmetric profile of the ST beam in the $x-z$ plane, this duality symmetry could be conserved under a spatial scaling transformation applied to the ST beam, i.e., $x \to \alpha x$, where $\alpha$ denotes the spatial magnification factor along the $x$-axis. Consequently, a generalized duality symmetry condition emerges, $\beta_2^* = \left(\frac{1}{\alpha^2}\right)\frac{-c}{n(\omega_0)\omega_0 v_g^2}$, with $1/\alpha^2$ acting as the scaling factor for the dispersion material to guarantee the 2D ST duality symmetry (see more details in Supplementary Text S1).

Notably, the propagation dynamics of the monochromatic spatial beam within the $x-y$ plane and the ST beam within the $x-z$ plane—as described by Eqs. (3) and (4), respectively—exhibit a remarkable parallelism under the 2D ST duality symmetry. To demonstrate and quantify this correspondence, we calculate the similarity $\gamma(z) = \left|\iint \mathrm{d}x\mathrm{d}\zeta\, \psi(x,y;z)\psi^*(x,\zeta;z)\right|^2 / \left(\iint \mathrm{d}x\mathrm{d}y\, |\psi(x,y;z)|^2 \iint \mathrm{d}x\mathrm{d}\zeta\, |\psi(x,\zeta;z)|^2\right)$ between a 2D spatial Gaussian beam and a 2D ST Gaussian wavepacket during propagation, where $*$ denotes complex conjugation. As shown in Figs. 1c and 1d, after a propagation distance of $20z_R$—where $z_R = \frac{\pi w_0^2}{\lambda_0}$ is the Rayleigh range, $w_0$ is the beam's spatial width, and $\lambda_0$ is the central wavelength—the similarity $\gamma(z)$ remains at unity ($\gamma = 1$) when $\beta_2 = \beta_2^*$. Conversely, $\gamma(z)$ decreases as the material dispersion deviates from $\beta_2 = \beta_2^*$.

ST Gaussian wavepackets are ST separable, suggesting they can be represented as the product of a 1D spatial Gaussian beam and a temporal Gaussian pulse; hence, breaking the 2D ST duality symmetry leads to independent broadening in the temporal and spatial domains. In contrast, for more generalized ST non-separable wavepackets, the various extents of 2D ST duality symmetry breaking fundamentally govern their unique evolutions. For instance, the STOV, as a solution to Eq. (4), inherently maintains its shape during propagating in a dielectric with $\beta_2 = \beta_2^*$. In free space, however, it undergoes a time-symmetrical evolution, behaving as if driven by a normal dispersion of $\beta_2^{\text{int}} = -\beta_2^*$, where $\beta_2^{\text{int}}$ represents the recently confirmed intrinsic dispersion of an STOV [26] (see more details in Supplementary Text S2 and Supplementary Fig. S1). On the contrary, the ST light sheet, as a diffraction-free solution to Eq. (4) under broken ST duality symmetry ($\beta_2 = 0$) [19], exhibits propagation invariance in free space but experiences mode evolution in the presence of material dispersion (see more details in Supplementary Text S3 and Supplementary Fig. S2).

## Arbitrary 2D ST wavepacket generator

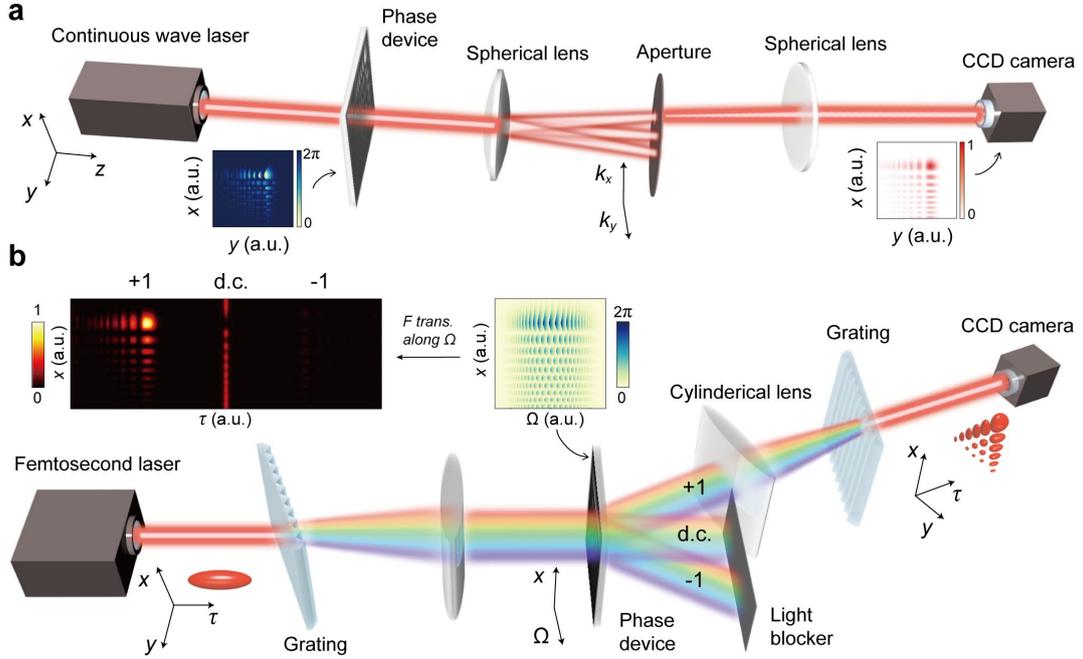

**Fig. 2 | ST complex-amplitude modulation. a,** In the conventional light shaping process (illustrated with a 2D spatial Airy beam), a monochromatic beam is first modulated in the $x - y$ domain by a phase device. Subsequently, the target light field, which is present in the +1 order diffracted field, could be extracted out in the spatial frequency $k_x - k_y$ domain by spatially filtering. Insets: Phase loaded in the phase device (left) and the resulting target light field (right) in the $x - y$ domain. **b,** Concept of ST complex-amplitude modulation (illustrated with a 2D ST Airy wavepacket). The temporal frequency spectrum of a pulsed Gaussian beam is broadened by a diffraction grating before encountering a phase device that imparts a wavefront modulation with phase $\varphi(x, \Omega)$. A light blocker is used to exclude all diffraction orders except for the +1 order, thereby extracting out the target light field $\tilde{\psi}_{\text{tar}}(x, \Omega)$, which could be then transformed into the target ST wavepacket in the $x - \tau$ domain by another grating. Insets: Phase loaded in the $x - \Omega$ domain (right) and the diffracted field after a Fourier transformation along $\Omega$ (left), where the +1 order corresponds to the target ST wavepacket.

A range of approaches for generating ST structured light fields has emerged, with most relying on ST Fourier pulse shaping for experimental designs [15–19, 21–24, 26, 28, 29], while a few theoretical works suggested employing micro/nanodevices with tailored transmission functions [30–32]. Notably, a shared constraint among these methodologies is their suitability for limited ST beam types. The principle of 2D ST duality, which showcases the mathematical congruence between spatial-structured light and ST beams, indicates the potential to generate arbitrary ST wavepackets extending from the conventional light shaping process. As shown in Fig. 2a, spatial-structured light is typically generated by modulating incident light with a phase device in the $x-y$ domain (illustrated with a 2D spatial Airy beam) [33]. The target light field, encoded in the +1 order of the diffraction field, is then extracted in the $k_x - k_y$ domain using a spatial filter composed of two spherical lenses and an aperture [33].

Analogously, modulating an incident pulse in the $x - \tau$ plane and extracting the target ST beam from the diffracted field would be a straightforward extension of the above process. However, due to the limited response time and bandwidth of current phase modulation devices, direct $x - \tau$ modulation remains significantly challenging. Instead, we propose here an ST complex-amplitude modulation capable of generating arbitrary ST beams, realized by modifying the conventional pulse shaper [34] that incorporates a phase device between a pair of diffraction gratings in a $4f$ system (see Fig. 2b). In our arrangement (illustrated with a 2D ST Airy wavepacket), the first diffraction grating transforms the incident pulse $\psi(x, \tau; z_0)$ into a light field $\tilde{\psi}(x, \Omega; z_0) = \int d\Omega \psi(x, \tau; z_0) \exp(i\Omega \cdot \tau)$ via a temporal Fourier transformation in the $z = z_0$ plane. After being modulated in the $x - \Omega$ domain by the phase device, the +1 order of the diffracted field carries the desired complex-amplitude $\tilde{\psi}_{\text{tar}}(x, \Omega)$ at $z = 0$, which could be extracted out by spatially excluding all other diffraction orders and then reconverted into the $x - \tau$ domain by the second diffraction grating (see Fig. 2b).

Our main challenge lies in generating the required field $\tilde{\psi}_{\text{tar}}(x, \Omega)$ inside the pulse shaper. Inspired by the spatial complex-amplitude phase-only holograms [35], we assume that the phase loaded in the phase device could be expressed as:

$$\varphi(x, \Omega) = \exp\left(iM(x, \Omega) \bmod \left(N(x, \Omega) + \frac{2\pi\Omega}{\Lambda}, 2\pi\right)\right), \quad (5)$$

where $0 \leq M(x, \Omega) \leq 1$ is the phase scaling coefficient, $N(x, \Omega)$ is determined by $\tilde{\psi}_{\text{tar}}(x, \Omega)$, and $\Lambda$ is the period of a 1D grating in the temporal frequency $\Omega$-axis. After a Taylor-Fourier expansion, one can see that the +1 order diffracted field after being modulated by $\varphi(x, \Omega)$ is equal to $-\text{sinc}(\pi M - \pi) \exp(i(N + \pi M))$. By letting $-\text{sinc}(\pi M - \pi) = |\tilde{\psi}_{\text{tar}}|$ and $N + \pi M = \tan^{-1}\left(\frac{\text{Im}(\tilde{\psi}_{\text{tar}})}{\text{Re}(\tilde{\psi}_{\text{tar}})}\right)$, we obtain:

$$M = 1 + \frac{1}{\pi}\text{sinc}^{-1}(|\tilde{\psi}_{\text{tar}}|), \quad N = \tan^{-1}\left(\frac{\text{Im}(\tilde{\psi}_{\text{tar}})}{\text{Re}(\tilde{\psi}_{\text{tar}})}\right) - \pi M, \quad (6)$$

where $\text{sinc}^{-1}(\cdot)$ stands for the inverse function of $\text{sinc} = \frac{\sin x}{x}$ (see more details in Supplementary Text S4). Since the temporal frequency $\Omega$ is expanded along the $y$-direction within the pulse shaper, one can effectively select only the +1 order diffracted field $\tilde{\psi}_{\text{tar}}$ through spatially filtering inside the pulse shaper (see Fig. 2b). Notably, the ST complex-amplitude modulation is performed in the $x - \Omega$ domain rather than the $k_x - \Omega$ domain, subtly hinting at the broken 2D ST duality symmetry in free space ($\beta_2 = 0$). Theoretically, our approach applies to arbitrary scalar ST light fields with fidelity just limited by the parameters of the phase device, such as the fill factor, reflectivity, and resolution.

## Single-step generation of a 2D ST Airy wavepacket

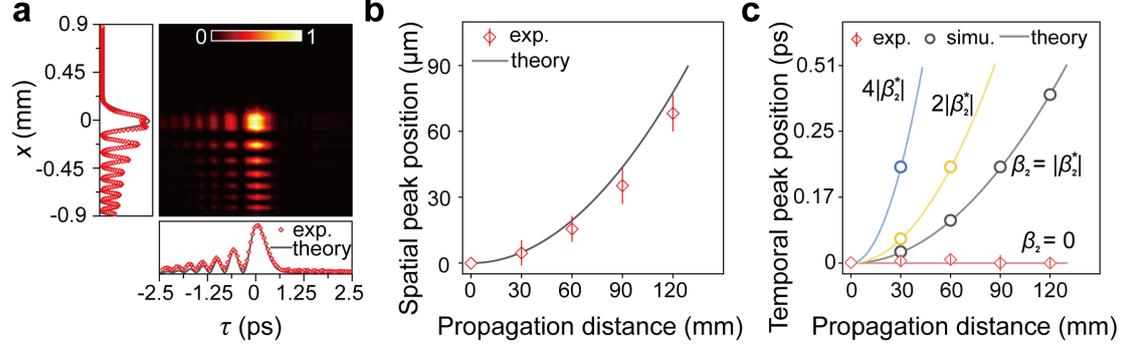

**Fig. 3 | Theoretical and experimental results for the generated 2D ST Airy wavepacket. a**, Intensity distribution of the generated 2D Airy wavepacket in the $x - \tau$ plane and its projections along the $x$- and $\tau$-axes. **b**, The acceleration effect of the generated 2D Airy wavepacket in the $x$-axis. **c**, Same as **b** but in the $\tau$-axis. Note that due to the breaking of 2D ST duality symmetry in free space ($\beta_2 = 0$), no temporal acceleration is observed for the Airy wavepacket; simulation results indicate that the temporal acceleration of the Airy wavepacket can be controlled by material dispersion, especially when $\beta_2 = |\beta_2^*|$, at which point its temporal acceleration profile will resemble the spatial acceleration profile.

To validate our proposed ST complex-amplitude modulation strategy, we first generate the previously investigated 2D ST Airy wavepacket that could be expressed as [36–38]:

$$I(x, \tau; z) = I_0 \text{Ai}^2\left(\varepsilon_x \frac{x}{x_0}\right) \text{Ai}^2\left(\varepsilon_\tau \frac{\tau}{\tau_0} - \frac{\beta_2^2 z^2}{4\tau_0^4}\right), \tag{7}$$

where $I_0$ is a constant, $\text{Ai}(\cdot)$ is the Airy function, $\varepsilon_x$, $\varepsilon_\tau = \pm 1$ determine the direction of the Airy function envelope, and $x_0$ ($\tau_0$) determines the spatial (temporal) width of the wavepacket. In the experiment, we start with femtosecond pulses having a central wavelength of ~800 nm and pulse duration of ~35 fs and use a commercial phase-only spatial light modulator (SLM) as the phase device, wherein the loaded phase is shown in Fig. 2b (see more details regarding the phase design in Supplementary Fig. S3). The generated Airy wavepacket ($x_0 \approx 90$ μm and $\tau_0 \approx 250$ fs) is measured by the Mach-Zehnder scanning interferometry [39, 40] with a ~80 fs reference pulse. Figure 3a shows the profile of the generated ST Airy wavepacket at $z = 0$, whose projections along the spatial ($\tau = 0$) and temporal ($x = 0$) axes resemble a 1D spatial Airy beam (with a main lobe of width of ~160 μm) and a temporal Airy pulse (with a main lobe of duration of ~440 fs), respectively.

It is evident that the ST Airy wavepacket freely accelerates following a parabolic trajectory in the spatial dimension (Fig. 3b), while its temporal peak remains stationary (Fig. 3c). This can be explained by the absence of dispersion in free space ($\beta_2 = 0$) breaking the 2D ST duality symmetry, which prevents the ST Airy wavepacket from behaving like a conventional spatial Airy beam. Additionally, due to the ST separability of the ST Airy wavepacket, one can disregard the sign of the material dispersion $\beta_2$ [36]. As shown in Fig. 3c, the simulated temporal acceleration of the Airy wavepacket at $\beta_2 = |\beta_2^*|$, $2|\beta_2^*|$, $4|\beta_2^*|$ aligns with the theoretical expectation of $\Delta\tau = \frac{\lambda_0^2 z^2}{16\pi^2 c^4 \tau_0^3}\left(\frac{\beta_2}{\beta_2^*}\right)^2$, where $\Delta\tau$ is the temporal peak position relative to the point $z = 0$ and $\lambda_0$ is the central wavelength (see more details in Supplementary Text S5). Interestingly, at $\beta_2 = |\beta_2^*|$, the ST Airy wavepacket demonstrates comparable parabolic acceleration in both spatial and temporal dimensions, with only minor deviations arising from its imperfect symmetry in the $x - z$ plane. By integrating the generation of 2D ST Airy wavepackets into a single-step ST modulation process, our approach offers a more economical and compact setup than separate spatial and temporal modulations in previous studies [37, 38].

## Generation of STOVs with a record-high fidelity

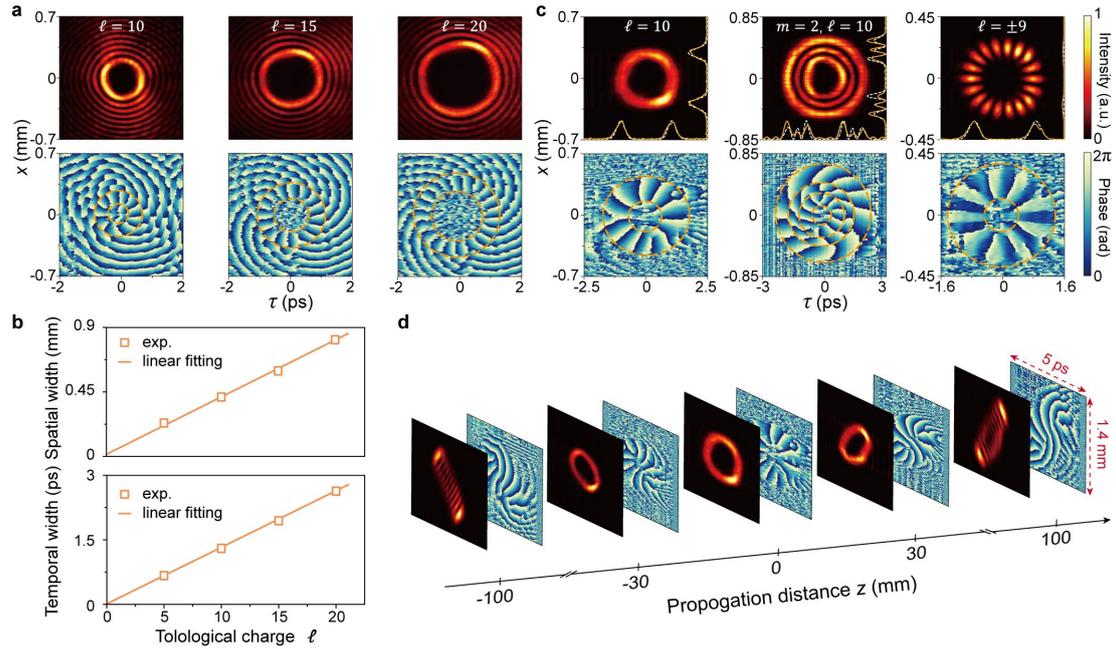

**Fig. 4 | Theoretical and experimental results for generated STOVs. a,** Reconstructed intensities and phases of generated STB vortices with topological charges of $\ell = 10$, 15, and 20. **b,** Dependence of spatial (temporal) diameter on the topological charges of generated STB vortices. **c,** Same as **a** but for generated STLG vortices with radial and angular mode numbers of $[m = 0, \ell = 10]$, $[m = 2, \ell = 10]$, and $[m = 0, \ell = \pm 9]$, respectively. In **c**, the intensity distributions of different STLG vortices at $x = 0$ and $\tau = 0$ are plotted with yellow solid lines for experimental and white dashed for theoretical. In **a** and **c**, the yellow dashed lines in the reconstructed phases highlight the characteristic ST spiral phases of the generated STOVs in the $x - \tau$ plane, indicating the transverse OAM. **d,** The measured temporal symmetric evolution of the generated STLG vortex $[m = 0, \ell = 10]$ across a propagation distance of $\Delta z = 200$ mm.

We next demonstrate the advantages of the ST complex-amplitude modulation scheme by generating a variety of STOVs possessing the transverse OAM. While our previous study has successfully synthesized ST Bessel (STB) vortices with topological charges up to $\ell = 100$, these vortices present noticeable asymmetric distribution against theoretical expectations [26]. We define here the fidelity $F$ as the overlap integral of the target ST wavepacket $\psi_{\text{tar}}(x, \tau)$ and the field generated by our proposed strategy $\psi_{\text{ge}}(x, \tau)$:

$$F = \left|\frac{1}{N_F} \iint dxd\tau \, \psi_{\text{tar}}(x,\tau)\psi_{\text{ge}}^*(x,\tau)\right|^2, \tag{8}$$

$$N_F = \left[\iint dxd\tau \, |\psi_{\text{tar}}(x,\tau)|^2 \times \iint dxd\tau \, |\psi_{\text{tar}}(x,\tau)|^2\right]^{\frac{1}{2}},$$

where $N_F$ is the normalization constant. Remarkably, the theoretical fidelity of the generated STB vortices by our strategy is $F = 97.5\%$, nearly three times the previously reported result [26] (see more details in Supplementary Text S6 and Supplementary Fig. S4). Experimentally, we observe the intensity and phase distributions of STB vortices (with a spatial spectral width of $\Delta k_x \approx 111$ rad/mm and a wavelength width of $\Delta\lambda \approx 11.2$ nm) with topological charges of $\ell = 10$, 15, and 20 (verified by the reconstructed phase distributions), with no noticeable asymmetry (see Fig. 4a; see measured $\ell = 5$ STB vortex in the Supplementary Fig. S5). Additionally, the measured STB vortices' spatial and temporal widths show a linear relationship with topological charges (see Fig. 4b), consistent with the previous result [26].

In conventional spatial-structured light research, Laguerre-Gaussian (LG) vortices draw significant attention due to their dual controllable mode numbers—radial ($m$) and angular ($l$), offering a wealth of mode combinations [41, 42]. However, due to technical limitations, previously generated ST Laguerre-Gaussian (STLG) vortices suffered from additional mode degeneration and were restricted to lower orders, such as $m = 0$ and $\ell = 1$ or 2 [21–24]. Utilizing ST complex-amplitude modulation strategy, we generate an STLG vortex ($\Delta k_x \approx 94$ rad/mm and $\Delta\lambda \approx 9$ nm) with $\ell = 10$ and a theoretical fidelity of $F = 98.4\%$ (Fig. 4c). We also demonstrate an STLG vortex with both non-zero radial and angular mode numbers [$m = 2$, $\ell = 10$], and an ST petal field formed by the superposition of two STLG vortices with different topological charges of $\ell = \pm 9$ (Fig. 4c). Moreover, we experimentally observe the temporal symmetric evolution of STLG vortex across a propagation distance of $\Delta z = 200$ mm (Fig. 4d), which is consistent with theoretical predictions [21, 43], confirming the effectiveness of our scheme in preventing the extra mode degeneration observed in earlier experiment studies.

**Towards more ST beams**

Notably, the 2D ST duality principle allows us to readily access a vast array of novel ST wavepackets, thereby significantly enriching the landscape of structured light. More specifically, the solutions to Eq. (3), which encompass all scalar spatial-structured light [44, 45], establish an exact one-to-one correspondence with the ST wavepacket solutions to Eq. (4). A noteworthy manifestation of this correspondence is the ST Hermite-Gaussian (STHG) wavepacket, which is a direct a solution to Eq. (4) (see more details in Supplementary Text S8). Figure 5a shows the experimentally generated STHG wavepacket $\text{HG}_{6,6}$ ($\Delta k_x \approx 96$ rad/mm and $\Delta\lambda \approx 10$ nm) represented as an ST Gaussian profile times Hermite polynomials of order $l = 6$ in the $x$-axis and order $m = 6$ in the $\tau$-axis. Thus far, the beams we have demonstrated have been in the ST plane as a product of functions in Cartesian $(x, \tau)$ coordinates. Expanding upon this concept, when we shift our perspective to elliptical coordinates, we naturally discover the ST Ince-Gaussian (STIG) and ST Mathieu-Gaussian (STMG) wavepackets (see more details in Supplementary Texts S9 and S10).

Experimentally, we generate two even and odd modes of STIG wavepackets $\text{IG}_{9,5}^e$ and $\text{IG}_{11,7}^o$ ($\Delta k_x \approx 96$ rad/mm and $\Delta\lambda \approx 10.6$ nm) with the ellipticity parameter $\epsilon_{IG} = 2$ respectively, with Ince polynomials replacing the Hermite polynomials of the STHG wavepacket (see Figs. 5b and 5c). We also generate two even and odd modes of different STMG wavepackets $MG_3^o$ and $MG_9^e$ ($\Delta k_x \approx 100$ rad/mm and $\Delta\lambda \approx 10.8$ nm) with the ellipticity parameter $q_{MG} = 27$ respectively, represented by Mathieu functions in the elliptic ST coordinates (see Figs. 5d and 5e). According to the principle of 2D ST duality, these beams exhibit propagation characteristics consistent with their spatial counterparts when the material dispersion satisfies $\beta_2 = \beta_2^*$, whereas spontaneous evolution occurs during their propagation in free space. Utilizing a charge-coupled device (CCD) camera, we directly measure the integrated intensity distributions $I(x;z) = \int d\tau |\psi(x,\tau;z_0)|^2$ of these ST wavepackets, over propagation distances of $10z_R$, where $z_R$ is the Rayleigh range of various Gaussian beams with spatial widths equal to those of the central peaks (bright or dark) in these ST wavepackets (see Fig. 5f). As can be seen, the STMG wavepackets still possess spatial quasi-non-diffraction characteristics, while the STHG and STIG wavepackets lack such features, which is consistent with the theoretical predictions. Moreover, the theoretical fidelity of the generated STHG, STIG, and STMG wavepackets are 97.7%, 98.1%, and 98.2%, respectively. These novel ST beams we demonstrated here may lead to innovative applications in optical manipulation and communications.

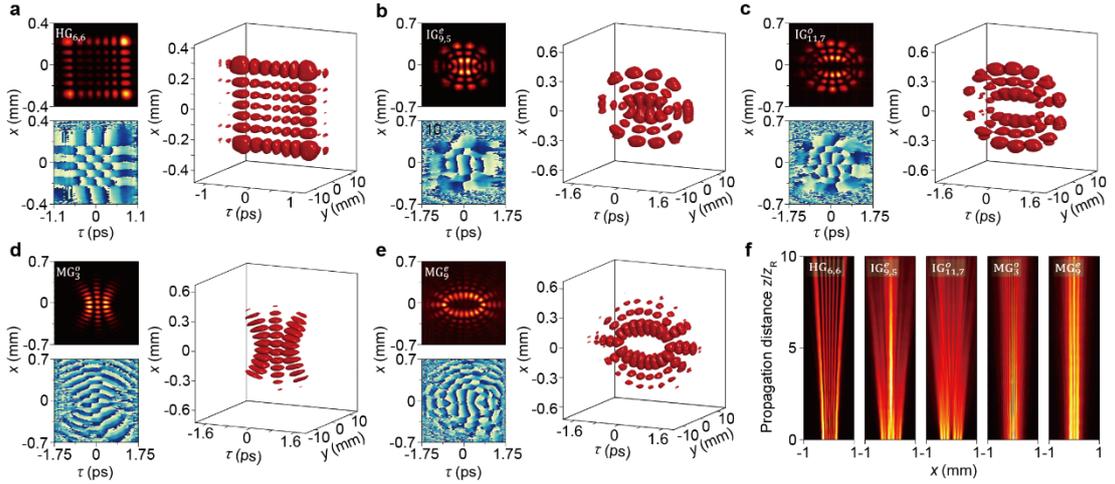

**Fig. 5 | Exploring ST beams harnessing the concept of 2D ST duality. a,** Reconstructed intensity, phase, and 3D profile of the generated STHG wavepacket with order $l = 6$ in the $x$-axis and order $m = 6$ in the $\tau$-axis. **b, c,** Same as **a** but for two generated STIG wavepackets $IG_{9,5}^e$ and $IG_{11,7}^o$ with an ellipticity parameter of $\epsilon_{IG} = 2$. **d, e,** Same as **a** but for two generated STMG wavepackets $MG_3^o$ and $MG_9^e$ with an ellipticity parameter of $q_{MG} = 27$. **f,** Measured integrated intensity distributions $I(x;z)$ for the generated ST wavepackets shown in **a**–**e**, where the $z_R$ is the Rayleigh range of different Gaussian beams having the same spatial size as the central peak widths (bright or dark) within these wavepackets.

## Discussion

We present a framework that leverages 2D ST duality to enrich the understanding and manipulation of ST wavepackets. This framework establishes a fundamental connection between the well-established realm of spatial-structured light and the exploratory terrain of ST beam research, rooted in the inherent duality in the paraxial wave equations governing these two beam types. Through precise engineering of the material dispersion, ST beams can be modulated to emulate or deviate from the evolution dynamics of their spatial counterparts. Concurrently, drawing upon principles from 2D ST duality, we introduce ST complex-amplitude modulation through an analogy with conventional spatial beam shaping techniques, enabling the generation of arbitrary ST wavepackets with an ultra-high fidelity exceeding 97%. Moreover, this framework uncovers an exact duality-based correspondence between ST beams and their spatial counterparts, offering new avenues for exploring and discovering ST wavepackets.

The potential of 2D ST duality has not been fully explored, and further harnessing the extensive research on spatial-structured light may significantly propel the study of ST beams forward. For instance, the established one-to-one correlation between scalar spatial-structured light and ST beams promises a wealth of as-yet-undiscovered ST wavepackets [44, 45]. By replacing the SLM in our methodology with geometric phase elements such as liquid crystals [46] or metamaterials [47, 48], we expect spin-dependent or time-varying polarized ST wavepackets to be generated by introducing polarization manipulation. While ST beams typically demonstrate a uniform distribution along the $y$-axis, their $y$-distributions can be further modulated by incorporating an additional spatial beam-shaping process [28, 49, 50], enabling the on-demand synthesis of electromagnetic structures in a higher-dimensional ST domain. Furthermore, 2D ST duality suggests the feasibility of studying ST beams within the $x - y$ domain (see more details in Supplementary Text S11 and Supplementary Fig. S6), which could significantly reduce the reliance on short-pulse light sources and complex interferometry in studying these beams. In summary, our findings bridge the gap between the established investigation of spatial-structured light and ST beam research, and concurrently imply an underlying principle for exploring ST phenomena across a broader range of wave systems, such as acoustic [51, 52], electron [53, 54], and even matter waves.

## Data availability

The data that support the findings of this study are available from the corresponding author on reasonable request.

## Code availability

The codes that support the findings of this study are available from the corresponding author on reasonable request.

## Acknowledgements

W.C., A. Y., L. M., Z. W., J. Y., and Y. L. acknowledge the support of the National Key Research and Development Program of China (2022YFA1405000), the Natural Science Foundation of Jiangsu Province, Major Project (BK20212004), Basic Research Program of Jiangsu Province (BK20232040), and the National Natural Science Foundation of China (NSFC) (62205136 and 62375119). Young Elite Scientists Sponsorship Program by CAST (2022QNRC001).

## Author contributions

C.W. and Y. L. proposed the original idea. C.W. performed all experiments and some theoretical analysis. A. Y. performed all theoretical analysis and some experiments. Z. Z., L. M., Z. W., J. Y., and C. Q. contributed to the theoretical model and the experimental implementation. L. M., C. Q., and Y. L. guided the data analysis and supervised the project. All authors contributed to writing the manuscript.

## Competing interests

The authors declare no competing interests.